\documentclass{article}
\usepackage[utf8]{inputenc}
\usepackage[margin=1in]{geometry}
\usepackage{times}
\usepackage{helvet}
\usepackage{courier}
\usepackage{color}
\usepackage{tikz}
\usepackage{hyperref}
\usepackage{graphicx}
\usepackage{tabularx}
\usepackage{float}
\usepackage{natbib}
\usepackage{amsmath}
\usepackage[utf8]{inputenc}
\usepackage{booktabs}
\usepackage{setspace}
\begin{document}
%

\title{Friend or Foe: A Review and Synthesis of Computational Models of the Identity Labeling Problem}
\author{Kenneth Joseph* \qquad University at Buffalo, Computer Science \\ 
Jonathan Howard Morgan \qquad Fachhochschule Potsdam
 \\ \quad \\
 * Email: kjoseph@buffalo.edu \quad Address: 335 Davis Hall, Buffalo, NY
}
\maketitle
\begin{abstract}
We introduce the identity labeling problem - given an individual in a social situation, can we predict what identity(ies) they will be labeled with by someone else? This problem remains a theoretical gap and methodological challenge, evidenced by the fact that models of social-cognition often sidestep the issue by treating identities as already known.  We build on insights from existing models to develop a new framework, entitled Latent Cognitive Social Spaces, that can incorporate multiple social cues including sentiment information, socio-demographic characteristics, and institutional associations to estimate the most culturally expected identity. We apply our model to data collected in two vignette experiments, finding that it predicts identity labeling choices of participants with a mean absolute error of 10.9\%, a 100\% improvement over previous models based on parallel constraint satisfaction and affect control theory.
\end{abstract}

\section{Introduction}

Identities are the labels we use to categorize ourselves and others. Examples of identities include social roles, like ``doctor’’ and ``mother’’, group memberships like ``Democrat’’ and ``Yankees fan’’, and social categories like ``black people’’ and ``women’’ (Tajfel and Turner 1979; Smith-Lovin 2007; Owens, Robinson, and Smith-Lovin 2010). Both the ways we label (or categorize) ourselves and others with identities, and the ways others label us impact our behavior. Social psychological theories such as identity theory \citep{stryker_burke_2000}, identity control theory  \citep{burke2009identity}, and social identity theory \citep{hogg2016social} describe the various ways identities are fundamental to our sense of self. 

In order to explain how identity impacts behavior, however, these theories must also address how we categorize (label) ourselves and others with identities in the first place. Each of these research programs has addressed categorization processes in some form. For example, identity theory specifies the probability that a person will choose a particular identity in a given situation and their commitment to that identity as a function of previous interaction opportunities, with interaction opportunities varying depending on an individual's social network \citep{stryker_burke_2000}. 
Whereas, social identity theory scholars have tended to conceptualize identity selection as signaling group affiliation in an effort to minimize social ambiguity and risk \citep{hornsey2008social}. All of these theories acknowledge categorization is important, but none, that we are aware of, have developed a framework capable of evaluating which social signals are important in categorization or their relative importance across a variety of social situations. We introduce such a framework. 

To our knowledge, only two existent conceptual approaches, \emph{affect control theory (ACT)} \citep{heise_affect_1987} and \emph{parallel constraint satisfaction-based models (PCS models)} \citep{freeman_dynamic_2011,schroder_affective_2013}, are capable of modeling identity labeling computationally.  These theories generally indicate that a small number of theoretically motivated dimensions of social meaning can capture much of the variability in how we label others.  However, what dimensions and how they are operationalized vary in critical ways between the two models, and the ramifications of this difference in terms of their predictive power is unknown. 

To address this gap in understanding, the present work empirically compares predictions of ACT and the PCS model of \citet{freeman_dynamic_2011}, which we call \emph{PCS-FA}, with a focus on understanding how differences in their operationalizations lead to different predictions. To formalize this comparison, we first mathematically characterize the problem of creating a predictive model of categorization in a broad array of social situations. In addition, motivated by the complementary focii of ACT and PCS-FA, we introduce a new framework to model how individuals select identities for others, which we entitle \emph{Latent Cognitive Social Spaces (LCSS)}. LCSS extends the mathematics of ACT to incorporate the theoretically motivated dimensions of social meaning of both ACT and the PCS-FA model. We then instantiate a survey that allows us to compare the predictions of ACT, PCS-FA, and LCSS. We find that ACT and PCS-FA are better able to predict survey responses in expected patterns that underlie their complementary theoretical underpinnings, and that LCSS, as a combination of the two models, explains the data significantly better than either model on its own.

Our work makes two contributions to the literature on how we determine the identities of those around us in social situations. First, our work is the first that we are aware of to mathematically formalize the following question - given an individual in a social situation, can we predict what identity(ies) he or she will be labeled with by someone else? We call this the \emph{identity labeling problem}. The identity labeling problem can be considered both a subset and a superset of the well-established notion of the definition of the (social) situation \citep{thomas_child_1928}. The identity labeling problem is one aspect of the definition of the situation in that the latter is an umbrella term that encompasses a variety of other processes through which a social actor internalizes and organizes a social situation in her head  (e.g. the labeling of settings and behaviors) in order to engage in ``self-determined lines of action and interaction’’ \cite[][pg. 63]{ball_definition_1972}. The identity labeling problem is also a superset of the definition of the situation, however, in that it applies in settings where definition of the situation does not obviously extend, particularly with respect to decisions in written language.
The second contribution of our work is LCSS, a novel combination of the core underlying principles of ACT and PCS-FA.  Our combination presents a mathematical formalization for how to combine cognitive and affective signals within a single framework that has eluded both the ACT and PCS communities. 

Below, we first provide a formal definition of the identity labeling problem. We then give an overview of ACT and PCS models, including PCS-FA, and introduce LCSS.  Following these overviews, we detail our experimental setup, and how we operationalize ACT, PCS-FA, and LCSS for this specific experimental task. We then present our results. Finally, we close with a discussion of new avenues of research that may extend the new problem we have defined and the new framework we have proposed to address it. All materials necessary to replicate our work can be found at (GITHUB LINK REMOVED FOR BLIND REVIEW).

\section{The Identity Labeling Problem}\label{app_sec:formal}

The identity labeling problem can be formally stated as follows:
\begin{quote}
Given a ``labeler,'' $x$, and a person to label, $y$, in a context with a particular set of active environmental cues $r_t$ at time $t$, predict the set of identities $\mathbf{i_{y}}$ in the universe of all identities $I$ that $x$ will apply to $y$
\end{quote}

We can derive from the problem statement above a mathematical form for how an identity labeling model uses a \emph{meaning space}, $M$, and a \emph{matching function}, $\phi$, to select a set of labels for $x$ to apply to $y$. In doing so, it is reasonable to assume that an element of randomness is inherent in $x$'s decision. Because of this, the result of an identity labeling model is best given as a probability distribution, where all identities have some (perhaps very small) chance of being applied. Formally, the probability that $x$ labels $y$ with a randomly selected set of identities $\mathbf{i_a}$ given environmental cues $r$ can be expressed as:
\begin{equation} \label{eq:orig}
	p(i_y = \mathbf{i_a} | \phi_x, r_{tx}, M_x) 
\end{equation}

This expression says that every person, $x$, may a) have a different meaning space $M$, b) observe or choose to attend to a distinct set of environmental cues, $r_t$, and c) use this information in a distinct fashion from others ($\phi$).  In the present work, however, we assume that $\phi_x = \phi, M_x = M, r_{tx} = r \, \forall \, t, x$. In other words, for the experiments discussed here, we assume that all survey respondents have the same meanings of identities and environmental cues, engage with the same set of cues and that all respondents use these meanings and cues in the same way to label $y$. Such simplifications are made in concert with prior work. Both affect control theorists and cognitive psychologists make similar assumptions in their existing identity labeling models, and so for the purposes of comparing models we make the same assumptions.

Having defined a probability distribution of interest, we now can explicitly define how a generic identity labeling model might leverage a particular $\phi$, $M$ and $r_t$ to define this probability distribution over identities.  Assume that $\phi$ is a function with three parameters. First, $\phi$ takes some identity, $i_a$, to be \emph{scored}. In our experimental design below, for example, we will score two identities, given to participants in questions like those posed in Figure~\ref{fig:example_q} (e.g. ``college student'' in Figure~\ref{fig:example_q}A).  Second, it takes in the meaning space of the theory, $M$. In ACT, for example, the meaning space constitutes the assumed positions of identities and traits in a three-dimensional affective meaning space. Finally $\phi$ takes the active environmental cues, $r$, from the current situation. Such cues might include, for example, the setting, or the behavior that $y$ is taking.   The function $\phi(i_a, M,r)$ then assigns a score to the given identity for how likely it is to be applied to $y$ in the current situation. 

A scoring function on each element of a set can be turned into a probability distribution by way of a \emph{discrete choice model} (DCM) \citep{mcfadden_econometric_1980}.  We can, therefore, formally define $p(i_y = \mathbf{i_a} | \phi, r, M)$  via a DCM, leading to the following:

\begin{equation}\label{eq:choice}
	p(i_y = i_a |M, r, \phi) = 
		\frac{ e^{\phi(i_a, M, r)}}
		     {\sum_{j \in I} e^{\phi(i_j, M, r)}}
\end{equation}

An identity labeling model can therefore be defined entirely by the way it defines the scoring function $\phi(i_a, M, r)$ and its parameters.  This is important, because it allows us to translate non-probabilistic predictive models, including ACT and PCS-FA, into a consistent, probabilistic framework that is 1) comparable and 2) can generalize to future identity labeling models.  

\section{Affect Control Theory}\label{app_sec:act}

\subsection{Overview}

ACT is a sociological social psychology model of identity and action. It has three core components: the measurement model, the impression change model, and the control model.   The measurement model defines an (affective) meaning space of identities, modifiers such as old or young, and behaviors. In ACT, each concept is defined by a three-dimensional affective, or sentiment, profile, on the Evaluative, Potency, and Activity dimensions. ACT's second component, the impression change model, specifies how the meanings of concepts change as a result of being combined, either modifying a concept to adjust for individual-level differences such as an old women \citep{averett1987modified} or in the context of a social event \citep{heise_expressive_2007}. Social events are conceptualized as an actor directing a behavior towards an object person. The impression change models are linear models that predict the situational meaning generated by combining the out-of-context or fundamental meanings of an identity and a modifying concept (e.g., an old man), or of an actor, behavior, and object person (e.g., a mother hugs a baby). Researchers estimate these equations using event stimuli that systematically vary concept combinations with respect to their evaluation, potency, and activity to create a theoretically comprehensive event space. ACT’s third component, the control model, predicts responses to events by optimizing for the behavior that will minimize the difference between the fundamental or widely held cultural meanings of the actors, behaviors, and object persons and the situational meanings resulting from the event. The intuition is that humans act in ways that maintain their identity and the identities of those with whom they are interacting whenever possible \citep{heise_expressive_2007}. 

The impression change model and the control model of ACT provide a mechanism to answer the identity labeling problem. 
Specifically, given that one identity and behavior are known, the label of the other identity can be derived from ACT. This provides a simple model of the identity labeling problem – given some partial information about a social event, we can identify unlabeled individuals participating in it. Recent advances have developed further this model of the identity labeling problem. In particular, a recent Bayesian reformulation of affect control theory, called BayesACT \citep{hoey_bayesian_2013,hoey_affect_2016,schroder_modeling_2016}, can infer the identity of an individual that engages in a social event when all identities are unknown. For example, given a social event where we observe one person instructing another, we can over the course of multiple events predict using BayesACT that the most likely identities of these two people are ``teacher’’ and ``student.’’ 

(Bayes)ACT is the only quantitative sociological model, to our knowledge, where the identity labeling problem is formalized and generalizable predictions are created. 
However, while BayesACT can be used to address the identity labeling problem, it was not developed for this purpose. Consequently, aspects of the theory make it unreliable when predicting identity labels. These difficulties are made clear in the following simulation example from \citet[][pg.846]{schroder_modeling_2016}: ``...both agents have developed the shared belief that one of them (agent A) is a ‘conservative’ while the other (agent B) is a ‘great grandmother’.'' While affectively, this shared belief may make sense given the behaviors these agents engage in towards each other, intuition tells us that the identity pairing of ``conservative'' and ``great grandmother'' is unlikely in a real world scenario, or at least, less likely than other potential combinations. For example, when we see a conservative, we typically expect to see a liberal, and where we identify a great grandmother, a ``great grandchild''. Improbable combinations such as the one just described result from relying exclusively on sentiments. In social interactions, people use sentiments, traits, and association information to make inferences about identities. 

This is because ACT lacks a mechanism for inferring identity information based on the types of semantic coherence that make great grandmothers and conservatives less likely to be discussed together than great-grandmothers and great-grandchildren. To address this, ACT scholars employ a semantic perspective when modeling events by defining simulations with respect to social settings \citep{heise_self_2010}. For example, if we know that an individual is in the ``school'' setting, ACT assumes that we will first use an filter that restricting labels for that individual to identities within the school setting, like principal, staff, teacher, and student. Although the theoretical relationship between institution and role-set is sound \citep{merton_role-set:_1957}, the operationalization of institutions within ACT has traditionally been theoretically unsatisfying because it does not address the questions about institutional boundaries, overlap, or how to conceptualize non-institutional identities such as strangers, loners, and friends. \citet{heiseCulturalMeaningsSocial2019}, in his recent book, addresses many of these theoretical concerns, and shows consistent patterns in the emergence of roles within institutions. The present work extends those efforts in a novel way to incorporate institutional affiliations of identities into an ACT-like model in a parsimonious way that also connects to the PCS framework.

In addition to questions about how to conceptualize institutions with respect to affective meanings, this understanding of semantic meanings also does not account for the fact that some semantic relationships are stronger than others. For example, the semantic association between ``student'' and ``teacher'' is probably stronger than the one between ``student'' and ``school administrator.'' In addition, the sequential processes suggested by the institutional filtering approach is inconsistent with what we know about cognition and the types of analogical relationships implied by institutions \citep{thagard1990analog}. Indeed, our results strongly suggest that humans use both types of information simultaneously. For example, when assessing whether a doctor is more likely to punch a patient or a trespasser, respondents rely on both semantic institutional information about the role of doctors as healers, and on affective information about the behavior of punching. The sequential ordering suggested by the filtering approach privileges patient in this case because doctors primarily associate with patients, but patient is clearly wrong. Doctors that behave like doctors do not punch patients. 

(Bayes)ACT, thus, gives an important conceptual model of how affective and semantic association meanings might be combined to make predictions for a specific form of the identity labeling problem. Further, ACT provides a grounding for how we can connect a model of identity labeling to the behaviors that these labels produce. Nevertheless, the precise mechanisms currently implemented by the theory are not likely to be useful in making identity labeling predictions in many instances, for example when role pairs such as mother/child can occur in multiple institutions (e.g., the family and education) depending on the context and other identities involved. More generally, ACT does not incorporate (and hence has no measurements for) semantic representations, and the relationships between affective meanings and semantic features remains underspecified in the formal model. 

\subsection{Mathematics of ACT}

We here introduce the basic mathematical model of ACT. We focus on ACT, as BayesACT derives much of its mathematics from the original model. ACT details how social events can be used to infer labels for identities, and thus to solve the identity labeling problem. Here, we provide mathematical details on how this process works.

Affect control theory models how the impressions we have of people change as result of social events. Social events are conceptualized as consisting of actors directing behaviors toward object people (e.g., the mother hugs the baby), each event element has a widely held cultural meaning referred to as a fundamental sentiment ($f$). When these meanings combine, our impressions of the actor, behavior, and object person shift in response. We refer to these new meanings as transient meanings ($\tau$) because they are result of particular situation which, in itself, is unlikely to change our general impressions of mothers, hugging, or babies but does change how we perceive these particular instantiations of these people and this behavior. We define both $f$ and $\tau$ as vectors of length nine, one element each for the $E$valuative, $P$otency and $A$ctivity affective dimensions for the $a$ctor, $b$ehavior and $o$bject, i.e. for $f$, we have:
\begin{equation}\label{eq:fundamental}
	f = \begin{bmatrix} a_e & a_p & a_a & b_e & b_p & b_a & o_e & o_p & o_a  \end{bmatrix}
\end{equation}

ACT also allows us to model how individual traits and emotions influence our impression of an identity (e.g. a ``bad teacher'' is different than a ``teacher'').  Let us assume there was a modifier such as bad describing the state of an actor. In this case, the fundamental is as follows, where $v$ is some modifier and $Mod$ is the regression equation for modifiers specified in \citep{heise_expressive_2007}:

\begin{equation}\label{eq:fundamental_mod}
	f = \begin{bmatrix} Mod_{v}(a_e) & Mod_{v}(a_p) & Mod_{v}(a_a) & b_e & b_p & b_a & o_e & o_p & o_a  \end{bmatrix}
\end{equation}

ACT specifies an equation that determines the values of $\tau$ as a function of $f$. This equation can be characterized by the form $\tau=\mathcal{Z} \, g(f)$, where the value of $g(f)$ is a $k$x$1$ vector of covariates and the matrix $\mathcal{Z}$ is a $9$x$k$ matrix specifying 9 different sets of regression coefficients, one for each element of $\tau$.  The actual values of $g(f)$ and $\mathcal{Z}$ are estimated via regression using survey data; the reader is referred to \citep{morgan_distinguishing_2016} and \citep{averett1987modified} for details on this process. 

In the present work, we will assume that $g(f)$ and $\mathcal{Z}$ are given.  We can then compute the post-event transient as follows, where $\mathcal{Z}_x$ represents row $x$ of the coefficient matrix:
\begin{align}
	\tau &= \begin{bmatrix} \mathcal{Z}_{a_e}^T \cdot g(f) & \mathcal{Z}_{a_p}^T \cdot g(f) & \dots  & \mathcal{Z}_{o_a}^T \cdot g(f) \end{bmatrix} \nonumber
\end{align}

Given $f$ and $\tau$, we can compute the \emph{deflection} of a social event as the unnormalized Euclidean distance between the fundamental sentiments of the event elements and the post-event transients, where the importance of each affective dimension can potentially be weighted by some weight vector $w$. In practice, $w_j$ is generally set to 1 for all elements of the fundamental:
\begin{equation} \label{eq:def}
	deflection(f) =  \sum_j^9 w_j (f_j-\tau_j)^2 = \sum_j^9 w_j (f_j-\mathcal{Z}_j\cdot g(f))^2
\end{equation}

Deflection gives an idea of how expected a social event is. An event where deflection is high indicates that the event significantly changes the affective meanings of the actor, behavior and object. Because ACT expects these meanings to be relatively consistent over time, this implies that the described social event is unlikely. Similarly, a low deflection value signifies an event that ``makes sense.''

\subsection{ACT as an Identity Labeling Model}

Starting with Equation~\ref{eq:def}, we can use ACT as an identity labeling model.\footnote{For a discussion of how BayesACT, ACT's successor, can be similarly used, see Appendix~\ref{app:bayesact}.}  Specifically, Equation~\ref{eq:def} can be used as a $\phi$ function for the identity labeling problem in the case where an actor or object-person with an unknown identity enacts (receives) a behavior on an object-person or from an actor with a known identity. In the language of the identity labeling problem, where we require a function $\phi(i_a,M,r)$, we can say that $i_a$ is a potential label (to be scored) for the unlabeled actor (or object-person), $M$ is the culturally defined EPA space given by ACT, and $r$ is the behavior the actor (object-person) is engaging in (receiving) as well as the identity of the object-person (actor) that $i_a$ is acting (being acted) upon. In addition, $r$ can include modifier terms; for example, we use the names of individuals in our vignette experiments as modifiers. The only distinction to be made is that, because events with higher deflection suggest more \emph{un}expected events, and $\phi$ should be higher with identities that better fit the situation, we negate the deflection score in our construction of $\phi$. 

Let us assume we wish to identify an unlabeled actor. Then for each potential identity in $I$, $i_a$, that we wish to score, we have environmental conditions $r$ composed of a behavior, $b$ and object, $o$, and meanings derived from EPA space. This gives the following, where, e.g., $M_{x_e}$ represents the Evaluative dimension of entity $x$ and $\circ$ concatenates two vectors:
\begin{align} \label{eq:act_phi}
    \phi(i_a, M, r) &= - deflection( i_a \circ r  ) \nonumber \\ 
    \mathrm{where} & \quad i_a = \begin{bmatrix} M_{i_{a_e}} & M_{i_{a_p}} & M_{i_{a_a}} \end{bmatrix} \nonumber \\
     \mathrm{and} & \quad r = \begin{bmatrix} M_{b_e} & M_{b_p} & M_{b_a} & M_{o_e} & M_{o_p} & M_{o_a} \end{bmatrix}
\end{align}

By computing deflection across all identities, $i_a$, in $I$ and then substituting it into the discrete choice model in Equation~\ref{eq:choice}, we can construct the desired probability distribution for the identity labeling model.  This is exactly the procedure we employ here, using regression equations and EPA values defined by \cite{heise_expressive_2007} and provided in the code and data release for this article. The only exception to this are the EPA values of the Actor names in our first experimental task, for which no EPA values exist.  We discuss this point further below, where we explain how we adapt each general model to our experimental setting.

\section{Freeman and Ambady's Parallel Constraint Satisfaction Model}\label{app_sec:pcsfa}



\subsection{Overview}

Cognitive psychologists have taken traits and associations and used them to construct predictive models for identity labeling. Most existing models are built within a parallel constraint satisfaction (PCS) framework. In PCS models, semantic links (``constraints'') exist between nodes, which can be anything from identity labels to particular traits. Each node is ascribed a single attribute - a level of cognitive activation. Each node also starts with a base level of this activation value. Nodes can then be ``excited'' by external stimuli, at which point their base level of activation is increased by a set amount. Activation then flows through links in the system using a predetermined flow equation. Notably, links can either act as exciting links or inhibitory links. An exciting link between two schema means that activating one of the links will increase the activation of the other.  For example, an exciting link likely exists between facial features indicating old age and the identity grandmother. An inhibitory link means that increasing activation in one schema will decrease the activation of another.   For example, an inhibitory link might exist between these same older-aged facial features and the identity infant.  The existence of these inhibitory links is the primary benefit of PCS models over pure spreading activation models \citep{collins_spreading-activation_1975}.

An explicitly relational perspective on identity labeling makes it much easier to model how overlaps and intersections between identities, both implicit and explicit, may manifest during the labeling process \citep{penner_engendering_2013}. Further, varying the strength and valence of these interrelationships can help to explain how, for example, biracial women can be more likely to identify as multiracial than biracial men \citep{davenport_role_2016}. The use of a cognitive, relational perspective, therefore, allows PCS models the ability to provide an appealing explanation for a variety of well-known phenomenon in the identity labeling process, in particular the existence of intersectional identities \citep{crenshaw_demarginalizing_1989}.

Several PCS models have been developed that show how semantic relationships inform the identity labeling process \citep{schroder_affective_2013,kunda_forming_1996,freeman_dynamic_2011}. A representative model, the focus here, is that by Freeman and Ambady \citep{freeman_dynamic_2011}. We use PCS-FA as shorthand to discuss this model. In the PCS-FA model, nodes can be one of four types. Nodes at the cue level include visual and auditory features, such as an individual’s face. At the category level, nodes indicating social categories (that is, a form of identity) exist. At the stereotype level, nodes exist that represent traits, such as annoying. Finally, a higher-order level includes nodes such as prejudice and motivations. Connections exist across nodes at different levels, and activation starting anywhere in the network is passed through the network until stability is reached, at which point an identity (category) is probabilistically selected based on its level of activation. 

\cite{freeman_dynamic_2011} show how this general model of social cognition allows them to replicate several different empirical results about how individuals are labeled under different experimental conditions. These experiments display the generalizability of the PCS framework, but also expose two weaknesses.  First, while PCS models can be flexibly applied, their use for any specific setting requires one to fix a specific set of concepts and links between them using a combination of domain knowledge and results from prior empirical research. Thus, while the modeling framework of PCS-FA is itself general, any \emph{specific instantiation of the framework is not}.  For example, a plausible PCS model instantiated from the PCS-FA framework might be constructed with an \emph{inhibiting} link between the identities brother and sister, to represent the fact that no individual can be both a brother and a sister. Such a model, however could not also have an \emph{exciting} link between these two identities to represent the fact there is a role relationship between these two identities, because a link in a single model cannot be both exciting and inhibitory. Put more simply, in a single PCS model, brother cannot easily both excite and inhibit sister; we need two different PCS models that we could then apply to the two different settings where the implied gender and role connotations of these identities are important.

Second, modeling sentiment information in PCS models is difficult. PCS models often must rely on the implicit notion of an identity’s affective meaning rather than represent this explicitly in the model.  This is made most clear when considering how one might adapt behaviors into PCS models. Unlike cognitive associations we can represent between finite categorical variables, e.g. race and sex, identities can engage in a multitude of behaviors at any given moment. While in some cases behaviors might be tied to specific identities (e.g. teachers often ``teach''), ACT shows that any individual, with any identity, taking any behavior, provides information of use in identity labeling. This generalizable relationship between identity and behavior is not well-suited to a model with pre-defined link structures. Consequently, PCS models with static link values and a static set of nodes, such as the Freeman and Ambady model, are infeasible for use modeling how behaviors in a situation inform the identity labeling problem.

\cite{schroder_affective_2013} address this issue by developing a PCS model of behavioral priming that draws directly on ACT. In their model, the PCS model’s nodes are the Evaluation, Potency and Activity dimensions for different identity labels. Their model, therefore, re-purposes the PCS model within an affective framework, rather than explicitly combining cognitive and affective meanings as we seek to do here.  Consequently,  \cite{schroder_affective_2013} model leverages a PCS design to put forth an affective model, rather than a model that jointly considers affective and semantic information (e.g. their model, while it could in principle be extended, cannot represent a relationship between ``brother'' and ``sister'') because ties remain fixed.

 \subsection{PCS-FA as an Identity Labeling Model}
 
 PCS-FA assumes that the meaning space $M$ is described by some semantic network $N$. The network $N$ consists of a set of concepts, or vertices $V$, and a set of edges between these concepts $E$. Some of the vertices are the identities in $I$, others might be relevant to specific traits (e.g. brown hair), settings (e.g. courtroom), motivations, etc. A positive edge is placed between concepts that ``stimulate'' activation of the other - for example, the setting ``school'' stimulates the activation of the identity ``teacher.'' Similarly, a negative edge is placed between concepts that ``inhibit'' activation of the other - for example, the trait ``young'' might inhibit activation of the identity ``grandfather''.

Environmental cues, $r_t$, are given in the form of one or more concepts that are initially activated in a given situation. This activation then spreads through the network via a series of iteratively applied update equations across links. When two nodes share a positive edge, activation in one node increases the level of activation in the other. When two nodes share a negative link, activation in one node decreases activation in the other. We refer the reader to the appendix of \cite{freeman_dynamic_2011} for the exact specifications of these update equations. 

Here, we note only that after some time, the level of activation for each node settles to a stable value representing how cognitively activated that node is, given the original stimuli flowing through the network. These update equations, iteratively applied, thus specify the function $\phi$. PCS-FA then assumes that the likelihood of a particular identity $i_a$ being used to label an individual is proportional to the level of activation of $i_a$ versus all other identities in $I$.

\section{Latent Cognitive Social Spaces}\label{app_sec:COMB}

Given the complementary nature of ACT and PCS-FA, it seems reasonable that a model which combines these two may hold greater predictive power for the identity labeling problem. This arguement from a predictive standpoint is supplemented by growing neuroscientific evidence that emotion and cognitive mechanisms are thoroughly integrated in the brain \citep{lindquist_brain_2012,pessoa_cognitive-emotional_2013,barrett_theory_2017}.  However, how to actually create such a combination, mathematically and theoretically, is not immediately obvious. The second contribution of this work is to propose one approach to doing so.

The combined model we create, LCSS, extends affect control theory's mathematical structure to propose a more general dimensional model. We demonstrate in Section~\ref{sec:specific_comb} that the mathematical model we lay out allows us to make certain assumptions that make analyses of experimental data tractable without knowledge of all model parameters. Borrowing from ACT, we use the concept of deflection as a signal of situational tension to define a $\phi$ for the identity labeling problem. We, however, include more situational cues that trigger this tension when there is situational ambiguity. Our more general formulation of deflection is consistent with recent theoretical developments in the affect control theory literature such as  \citeauthor{heise_self_2010}'s (\citeyear{heise_self_2010}) cultural theories of people and \citeauthor{RN8}'s (\citeyear{RN8}) work on the management of denotative and connotative meanings. Nevertheless, the relationships between these cues is an active area of research that this research speaks to but does not seek to definitively resolve.

The primary difference between the ACT formulation of the identity labeling model and the LCSS formulation is, thus, in $M$, the set of requisite meanings.  Specifically, LCSS hypothesizes that in addition to sentiments, identities can be defined by social traits and associations as well. Our focus in this work is to introduce a framework for combining various kinds of situational cues, but we recognize that subsequent theoretical work will be necessary to demonstrate that deflection can, in fact, be viewed as a general response to the violation of cultural expectations across a variety of situational cues. Our primary theoretical contribution here is to provide the framework to pursue this kind of theoretical work.

We represent the sentiment, traits, and associations of an identity in LCSS with a position in a multi-dimensional latent space.  Note that unlike PCS-FA, LCSS proposes that associations and traits, while perhaps best modeled as a network, can be approximated by an adequate geometric space. This idea is not new; network scholars have for years either implicitly by visualization or explicitly via latent space statistical models represented networks spatially \citep{hoff_latent_2002}. Thus, while our model adopts a solely latent space approach, it is conceptually similar to simply combining network-based PCS and latent-space based ACT models of meaning.

The first three dimensions of this space are the Evaluation, Potency, and Activity sentiment dimensions. All identities and behaviors are given a position in this affective space. The second $|T|$ dimensions of this space define meaningful social traits upon which culturally shared definitions of identities exist. Importantly, we do not assume, as in ACT, that traits are denotative realizations of positions in affective space. Rather, when we refer to traits, we are referencing what are, in the sociological literature, commonly referred to as \emph{status characteristics}, which acquire meaning in combination with dimensions ACT can model \cite{markWhyNominalCharacteristics2009} but that have also been shown to be critical to model as dimensions of meaning on their own \cite{morgan_duality_2018}. Finally, the last $|K|$ dimensions of the meaning space for LCSS define dimensions along which we can reconstruct a network of semantic associations in a latent space. Again, our assumption is that these associations can be modeled as existing within a relatively small number of latent dimensions, which represent, for example, dominant institutional domains \cite{josephWhenWordEmbeddings2020a}. Collectively, this set of $3+|T|+|K|$ latent dimensions, and a set of specified positions of behaviors and identities within them, defines the meaning space $M$ for LCSS.

To incorporate these new meanings into a deflection model that we can use to score a potential identity, we must augment the existing deflection equation from ACT. First, the fundamental and transient vectors of ACT must be expanded. To do so, we will require some additional notation. As in Equation~\ref{eq:fundamental}, let $a$, $b$ and $o$ stand for the \emph{a}ctor, \emph{b}ehavior and \emph{o}bject in a social event, and let, e.g., $a_e$ stand for the evaluative dimension of the actor. Additionally, let $f$ and $\tau$ hold the same meanings as in Section~\ref{app_sec:act}. Finally, as implied above, let $t_0,t_1,...,t_T$ be a set of $T$ traits for which fundamental meanings exist, and $k_0,k_1,...,k_K$ be a set of $K$ associative dimensions for which fundamental meanings exist. Given this notation, we now define $f^*$ to represent the fundamental vector for LCSS:
\begin{align}
	f^* &= f \circ f_t \circ f_k \nonumber \\
	&\mathrm{where} \quad f_t =  \begin{bmatrix} a_{t_0} & a_{t_1} \dots & a_{t_T} & o_{t_0} &  o_{t_1} & \dots & o_{t_T} \end{bmatrix} \nonumber \\
	\quad & \mathrm{and}\quad  f_k = \begin{bmatrix} a_{k_0} & a_{k_1} \dots & a_{k_K} & o_{k_0} &  o_{k_1} & \dots & o_{k_K} \end{bmatrix} \nonumber
\end{align}

Having described a new fundamental, we now must characterize the way in which fundamental meanings are changed by an observed social interaction.  Mathematically, we will need to define the quantities $\mathcal{Z*}$ and $g^*(f^*)$, the matrix of regression parameters and coefficients, respectively, that are needed to determine transient meanings from fundamental meanings.  For reasons explained below, we also will define a new weight vector, $w^*$, that is of length $|f^*|$. Using these variables, Equation~\ref{eq:def_COMB} gives the deflection equation for LCSS:
\begin{equation} \label{eq:def_COMB}
	deflection_{LCSS}(f^*) =  \sum_j^{|f^*|} w^*_j(f^*_j-\tau^*_j)^2 = \sum_j^{|f^*|} w^*_j (f^*_j-\mathcal{Z^*}_j^T g^*(f^*))^2
\end{equation}

One can create an identity labeling model for LCSS as is done for ACT, simply using the updated deflection equation.
 


We have now argued that the two existing computational models of identity labeling, ACT and PCS, can be described as special cases of a more general framework. In addition, we have introduced LCSS, a new model of identity labeling that combines principles from ACT and the PCS model of Freeman and Ambady. We now turn to the description and evaluation of an experiment we design to compare the predictions of these three identity labeling models on a set of carefully constructed questions meant to illuminate the strengths and weaknesses of these models.  Below, we first introduce our Experimental Design (Section~\ref{sec:experiment}), then provide an operationalization of each model within the experimental setting defined (Section~\ref{sec:models}), and then describe results (Section~\ref{sec:results}).

\section{Experimental Design} \label{sec:experiment}

\begin{figure}
\centering
\begin{tabular}{c}
   \includegraphics[width=.9\linewidth]{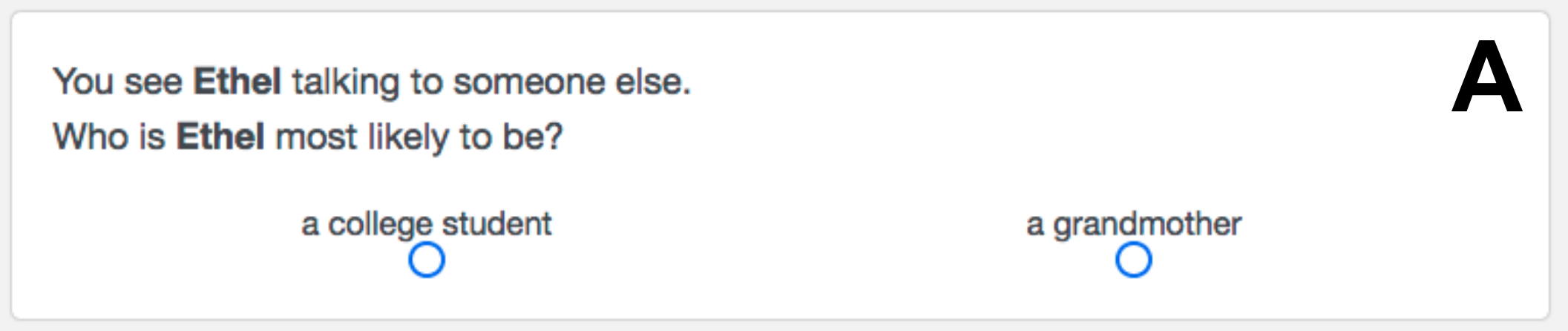}   \\
      \includegraphics[width=.9\linewidth]{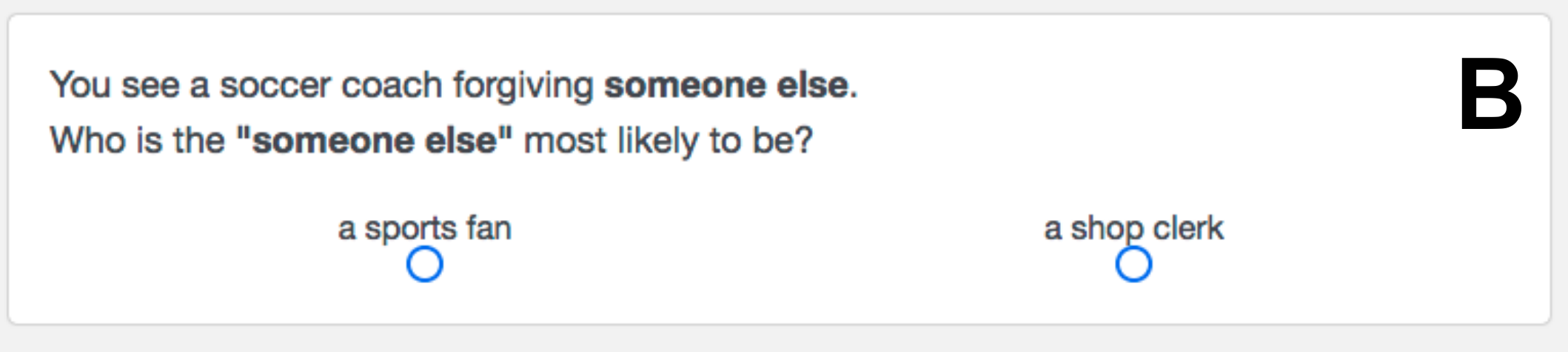}  
\end{tabular}
\caption{Example vignette questions from Study 1. A) Example for Task 1, which varied traits and sentiments.  B) Example  for Task 2, which varied associations and sentiments.}
\label{fig:example_q}
\end{figure}

\subsection{Overview}

In our experiment, we ask seventy-eight participants on Amazon’s Mechanical Turk to complete two tasks composed of a series of vignettes. Example vignettes are presented in Figure~\ref{fig:example_q}---we asked 16 questions in Task 1 similar to Figure~\ref{fig:example_q}A), and 24 questions in Task 2 similar to Figure~\ref{fig:example_q}B). Participants were presented with a situation in which an unlabeled (although potentially named) individual enacts a behavior on someone else (Task 1) or has a behavior enacted upon them (Task 2). Study participants were asked to label this person with one of two identities. 

Note that the questions posed in our experiment are simplistic instantiations of the more general identity labeling problem.  In the survey question in Figure~\ref{fig:example_q}a),  $x$ the survey respondent, and $y$ is the ``someone else'' to be labeled.  The universe of all identities $I$ is limited to the two answer choices given for each question, and $y$ is only ever labeled with a single identity $i_a$. Finally, across Task 1 and Task 2, the active environmental cues (i.e. the salient information given to the labeler) $r_t$ are two-fold. For both Tasks 1 and 2, a social event---where an actor enacts a behavior on an object-person ---provides salient cues about the sentiments of the people and the behavior. In Task 1, the social event contains trait-based and sentiment-based cues about $y$ in the form of a name. In Task 2, the actor additionally has varying degrees of semantic association to the answer choices. 

Vignettes in Task 1 varied traits and differences in affective deflection in a 2x2x2 design.
By traits, we mean socially relevant category memberships such as female or male, young or old. By affective deflection, we mean the expected surprise we expect respondents to have given the affective meanings of the identities and behaviors displayed. We varied vignettes with respect to either a high or low \emph{difference} in affective deflection, and used two trait signal conditions---age (old or young) and gender (female or male). We operationalized traits using gendered names with clear age connotations: Ethel (old/female), Harold (old/male), Brittany (young/female), and Johnny (young/male). In addition, we included a condition where no traits were implied, using the term ``someone'' in place of ``Ethel'' in Figure~\ref{fig:example_q}A.  This was used as a control condition where no trait information was provided.
In Task 2, we varied differences in affective deflection and associations in a 2x4 design. We varied vignettes with respect to a high or low difference in affective deflection between the potential answers, and high (role pairs such as soccer coach and soccer player), medium (identities in the same institution such as soccer coach and sports fan), low (identities that have no association such as soccer coach and shop clerk), or no (using ``someone'' instead of soccer coach in the question text) association with the identity in the question.  

We evaluated the extent to which four different models, described below, could explain the percentage of respondents who selected a given identity. For each question, we computed the absolute difference between the probability the model assigned to the identity and the actual percentage of respondents who selected it. When considering multiple questions, we compute the mean of this quantity, referred to \emph{mean absolute error (MAE)}. The four models we consider are ACT, Estimated PCS-FA, which is a PCS-FA estimated using the experimental data, Hand-coded PCS-FA, a PCS-FA with a static, a-priori parameter set, and LCSS.

\subsection{Participant Sample}
We recruited 80 participants on Amazon’s Mechanical Turk for the vignette experiments for Task 1 and Task 2.  These had greater than a 95\% completion rate and had completed over 1,000 HITs. Two participants were excluded due to their responses as to whether or not they were current residents of the United States, for a total of 78 respondents.  The sample was split evenly on Gender (39 male, 39 female), and the median age was 31.  We did not request information about the respondent's level of education or race.

We minimized the effects of sociodemographic characteristics on task performance by designing the tasks in such as way that any person socialized in the United States would be familiar with the identities, behaviors, and trait associations. For Task 1, we used culturally encoded names. These names are European names that while not common to all communities in the United States are generally recognized by most people. We tested this operationalization in pilot tests, as well as in the survey itself, by asking respondents to indicate what they would expect the age and gender of a person with this name to be.  For Task 2, we selected role identities that respondents are likely to have encountered in their daily lives such as doctors, patients, and paramedics performing behaviors respondents were likely to recognize such as forgiving, assisting, or punching. 

The survey was approved by the Institutional Review Board of \emph{REMOVED} University.

\subsection{Task 1}

\begin{table}[t]
    \centering
    \begin{tabular}{ l l  l l  l l l }
\midrule
\textbf{Actor} & \textbf{Behavior} & \textbf{Object} & \textbf{Answer 1} & \textbf{Answer 2} & \textbf{Trait Condition} & \textbf{Aff. Defl. Condition} \\ \midrule
Ethel & attacking & enemy & grandmother  & bully & Old,Female & High \\
someone & attacking & enemy &  grandmother & bully & No trait & High \\ \hline
Ethel & talking to & person &  grandmother & college student &Old,Female& Low \\ 
someone & talking to & person& grandmother  & college student & No trait & Low \\ \hline \hline
Harold  & attacking & enemy  & grandfather& bully & Old,Male& High \\ 
someone & attacking & enemy  & grandfather& bully & No trait & High \\ \hline
Harold & talking to & person & grandfather & college student & Old,Male& Low \\ 
someone & talking to & person & grandfather & college student &  No trait & Low \\ \hline \hline
Brittany & attacking & enemy & granddaughter & villain & Young,Female & High \\ 
someone & attacking & enemy & granddaughter & villain &  No trait & High \\ \hline
Brittany & talking to & person & grandson  & pharmacist & Young,Female & Low \\ 
someone & talking to & person & grandson  & pharmacist  & No trait & Low \\ \hline \hline
Johnny & attacking & enemy & granddaughter & villain & Young,Male & High \\
someone & attacking & enemy & granddaughter & villain & No trait & High \\ \hline
Johnny & talking to & person & grandson & pharmacist & Young,Male & Low \\
someone & talking to & person & grandson & pharmacist & No trait & Low \\ \hline \hline
    \end{tabular}
    \caption{Conditions for Task 1 questions  (note: Aff. Defl. stands for Difference in Affective Deflection)}
    \label{tab:study1_task1}
\end{table}

Figure~\ref{fig:example_q}a) provides a template for questions asked in Task 1.  In Table~\ref{tab:study1_task1}, we outline all of the questions asked along with the experimental conditions they belong to (the final two columns of the table). Each question asks the respondent to label either an individual with a name, or whom we called ``someone''.  This individual, the \emph{Actor}, is described as engaging in a \emph{Behavior} towards an \emph{Object}, as in ACT's social event model. Respondents are then given two choices for the identities of the Actor, and asked to select the one that fits best.  So, for example, the first row of the table would represent the question: ``You see Ethel attacking an enemy, is Ethel a bully or a grandmother?'' This question tests a condition where the implied traits of the name are Old \& Female, and there is a high degree of difference in affective deflection between the two potential identities given answers.

We use the name information to encode trait information, and consider four different trait characteristics - ``Ethel'' is used to depict the traits ``Old \& Female'', Harold for ``Old \& Male'', ``Brittany'' for ``Young \& Female'' and ``Johnny'' for ``Young \& Male.'' These individuals are engaged in a social event towards an Object who has an identity.\footnote{We included a manipulation check for the connection of these names to the associated ages by asking respondents how old they expected someone named ``Ethel'', ``Harold'', ``Brittany'' and ``Johnny'' to be. Our expectation was that Harold and Ethel would be of nearly equivalent age, and that this age would be appropriate for an elderly individual. We also expected that Johnny and Brittany would also be of a roughly equivalent age, and that this age would correspond roughly to the age of a child. On average, Ethel was expected to be around 73 years old; Harold around 60, Brittany around 29 and Johnny around 30. This suggests that the manipulation check worked better for the older identities than the younger identities, because they are much closer to the stages in the life-course we were intending. More importantly, however, is that there is a much larger lifecourse difference between age the participants reported for Harold and Ethel and those reported for Johnny and Brittany.}  The behavior taken, and the Object identity, are determined by the level of sentiment signal for the given condition. In the ``High'' (``Low'') Affective Deflection Condition, we use a social event that creates a salient (weak) difference in affective deflection between the two potential answers. The latter, ``someone'' questions serve as cases in which only sentiment signals are present, representing the ``No trait'' condition.

In all cases, Answer 1 is defined by matching the trait signal to an identity - for example, we match the identity grandson to the trait signal ``Young \& Male''.  Answer 2 maps to an identity predicted by ACT as the most likely identity given the cues provided. Answer 1 is distinct on the trait signal from Answer 2. For example, pharmacists may not be obviously male or female, but are likely to be implicitly characterized as being older than grandsons. Note that although we refer to Answer 1 and Answer 2, answer order was randomized for participants.

\subsection{Task 2}

\begin{table}[t]
    \centering
    \begin{tabular}{llllll }
    \hline
\textbf{Actor} & \textbf{Behavior} & \textbf{Answer 1} & \textbf{Answer 2} & \textbf{Aff. Defl. Condition} & \textbf{Associative Condition} \\ \hline
soccer coach & forgiving  & soccer player & shop clerk & Low & High (Role Pair) \\ \hline
doctor & assisting & patient & cousin & Low & High (Role Pair) \\ \hline
soccer coach& hurting & soccer player & goon & High & High (Role Pair)  \\ \hline
doctor & punching &  patient & trespasser & High & High (Role Pair)  \\ \hline
soccer coach & forgiving & sports fan & shop clerk & Low & Medium (Same Institution) \\ \hline
doctor & assisting &  paramedic &  cousin & Low & Medium (Same Institution) \\ \hline
soccer coach& hurting & sports fan & goon  & High & Medium (Same Institution) \\ \hline
doctor &  punching & paramedic & trespasser & High & Medium (Same Institution) \\ \hline
soccer coach& forgiving & goon & shop clerk & Low & Low (Different Institution) \\ \hline
doctor & assisting &trespasser & cousin & Low & Low (Different Institution) \\ \hline
soccer coach & hurting & shop clerk & goon & High & Low (Different Institution) \\ \hline
doctor & punching & cousin & trespasser & High & Low (Different Institution) \\ \hline
    \end{tabular}
    \caption{Conditions for Task 2 questions (note: Aff. Defl. stands for Difference in Affective Deflection)}
    \label{tab:study1_task2}
\end{table}

Figure~\ref{fig:example_q}b) provides a template for questions asked in Task 2, and Table~\ref{tab:study1_task2} outlines the specific questions asked and the associative and affective conditions being tested with each. In Figure \ref{fig:example_q}b), the Actor is a soccer coach, the Behavior is forgiving, Answer 1 is sports fan, and Answer 2 is shop clerk. Each question is drawn from one of two different affective deflection information conditions—high  or low —and one of three different associative information conditions--- high (association between the Actor and Answer 1), medium, or low. For each condition, we consider two different Actors, soccer coach and doctor, and construct social events these actors are engaged in based on the signal conditions.

The behavior the actor engages in is determined by the affective deflection condition. For the high sentiment deflection condition, we select strong, negative actions (hurting and punching), while for the low deflection condition we select relatively innocuous ones (forgiving and assisting).  As with Task 1, the high (low) deflection condition is then, by our construction, characterized by a salient (weak) difference in affective deflection between Answer 1 and Answer 2.  Answer 2 is always the most appropriate answer with respect to ACT, as determined by Interact, a simulation tool used by ACT scholars that implements the theoretical model \citep{heise_project_2001}. Answer 1 is, instead, determined by the associative signal condition. At high levels of associative signal, Answer 1 maps to a role pair for the actor (e.g. soccer players and soccer coaches, patients and doctors). At the medium level, Answer 1 is selected to be an identity in the same institution.  An institution is an organizing principle around which identities, behaviors and settings typically align \citep{heise_self_2010}.  For example, soccer coaches and sports fans are both members of the sports and leisure institution. At the low level, we simply select a random identity. In all cases, we attempt to choose identities that do not strongly imply trait-based information, and thus control for trait.

In addition to the twelve questions described in Table~\ref{tab:study1_task2}, we also include a set of 12 control questions. These control questions are identical to those described in Table~\ref{tab:study1_task2}, but instead of providing an actor identity they simply use the term someone (i.e. someone forgives someone else, who is that someone else?) These questions can be used to assess the affective model of ACT, absent any associative meanings. Thus, these questions imply a fourth condition, the ``None'' condition, for associative meanings, that we test with our survey questions.



\section{Specific Models Tested}\label{sec:models}

In order to evaluate ACT, PCS-FA, and LCSS using results from our experiment, we must address the fact that each model contains several potentially situation-specific parameters in either their meaning model, their matching function, or both. Below, we provide details on how these parameters are estimated to ensure a fair comparison across models.

Before continuing, one important note for the present work is that, in the special case of the identity labeling problem where the universe of possible labels, $I$, is a set of two options $i_a$ and $i_b$ (as in the survey data), we can further reduce Equation~\ref{eq:choice}. We show this below, where in the third step below we divide through by $e^{\phi(i_a, M, r)}$:
\begin{align}\label{eq:bin_case}
 p(i_y = i_a |M, r) &= 
		\frac{ e^{\phi(i_a, M, r)}}
		     {\sum_{j \in I} e^{\phi(i_j, M, r)}} \nonumber \\
		&= \frac{ e^{\phi(i_a, M, r)}}
		     {e^{\phi(i_a, M, r)} + e^{\phi(i_b, M, r)}} \nonumber = \frac{ 1}
		     {1 + \frac{e^{\phi(i_b, M, r)} }{e^{\phi(i_a, M, r)}}} \nonumber  \\
		&= \frac{ 1}
		     {1 + e^{\phi(i_b, M, r)-\phi(i_a, M, r)}}
\end{align}

The final line in Equation~\ref{eq:bin_case} is a binary logistic model where the activation function is defined by the difference between $\phi$ for $i_a$ and $i_b$. As we show below, given a particular form of $\phi$, this equation can be further simplified to the point where we can use standard logistic regression model to estimate parameters for PCS-FA and LCSS.

\subsection{ACT}

To apply ACT to our experiment, we required only that we gather EPA ratings for the four names in Task 1 of our experiment. These sentiment ratings were elicited using standard approaches in ACT, see \citep{heise_surveying_2010} for question details. For these, we provide our own survey respondents with traditional questions used to estimate EPA values, and use the mean values of these responses as EPA values for these concepts.  These values are, in turn, used as modifiers for the potential identities that respondents could select.

\subsection{PCS-FA}

To show how we construct an estimate of a PCS-FA model for our survey data, let us focus on Task 2. Here, the environmental cue consists of the already-labeled individual in the social event, which we will call $i_q$, and a behavior.  There are also two identities the respondent can choose from, let us call them $i_a$ and $i_b$. While in general, behaviors may have semantic relationships, we have selected behaviors for the present work that carry limited semantic attachments to the specified identities. Thus, we assume all behaviors we study here are isolates in the network $N$, and therefore cannot spread any activation. 

Consequently, we will focus only on the cue from $i_q$ when building our PCS-FA models for Task 2. Let us now assume that we have, in some way, constructed a $\phi$ and a network $N$; perhaps we use the original form of $\phi$ in \cite{freeman_dynamic_2011} and $N$ consists of a network linking $i_a$ and $i_b$, the two potential choices for our survey respondent, and $i_q$ all together with positive links of varying weights. We then input $i_q$ as our environmental cue $r_t$ and generate final activation rates for $i_a$ and $i_b$.

Let us call these final rates of activation $\beta_a$ for $i_a$ and $\beta_b$ for $i_b$. In the language of the identity labeling problem, these final activation rates represent the score outputted by the matching function of PCS-FA. Given this, and the result in Equation~\ref{eq:bin_case}, we have that:

\begin{equation}\label{eq:gen}
p(i_y = i_a |M, r) = p(i_y = i_a | N, i_q)
=  \frac{ 1} {1 + e^{\phi(i_b, N, i_q)-\phi(i_a, N, i_q)}} 
=  \frac{ 1} {1 + e^{\beta_b - \beta_a }}
\end{equation}

Let us now assume that $\beta$ is a vector of length $K$, where $K$ is the number of identities (4) in each of the two scenarios presented in Task 2.   These scenarios are distinguished by particular $i_q$ - in the first, $i_q$ is ``soccer coach'', in the second, it is ``doctor''.  

Let us now focus only on the first scenario. We define two binary vectors $a$ and $b$, which tell us what identities are provided as $i_a$ and $i_b$ for a particular question. Because there are four total identities in each scenario considered, $a$ and $b$ are of length four. We then assign an arbitrary position for each identity across any question answer in each scenario. So, for example, ``soccer player'' is assigned to position 1, ``shop clerk'' to position 2, etc. Given these assignments, when we ask, e.g., ``A soccer coach is assisting someone, is that someone a soccer player or a shop clerk'' (first row, first scenario in Table~\ref{tab:study1_task2}), then we define $a = \begin{bmatrix} 1 0 0 0 \end{bmatrix}$  and $b = \begin{bmatrix} 0 1 0 0 \end{bmatrix}$. A similar construction exists for all other questions provided. 

Using this formulation, we can extend Equation~\ref{eq:gen} to define the probability that a given survey respondent selects $i_a$ for all survey questions for this scenario:
\begin{equation} \label{eq:est_beta}
    p(i_y = i_a | M, r) = \frac{ 1} {1 + e^{\beta_b - \beta_a }} = \frac{1} {{1 + e^{\beta^Tb - \beta^Ta }}} = \frac{1} {{1 + e^{\beta^T(b - a) }}}
\end{equation}

Here, a given index of $\beta$ represents the score produced from the matching function of a PCS-FA model given the cue $i_a$ for a particular identity.  If we can determine sensible values for this final score, we need not worry about the exact specification of $\phi$ and $N$ for our survey data. Note that in our case, where a small number of vignettes are considered, there is little difference between specifying this metric by hand and defining a meaning space (semantic network) $N$ by hand, because we can always construct a meaning space to generate any values of $\beta$ that we might deem reasonable anyway.

More importantly, by assuming we are only interested in these final values, we can use our data to learn the best possible values for $\beta$  with respect to their consistency with the survey data.  Thus, we can develop scores that we can assume were generated under the best possible meaning space for our data under the assumption that traits and associations, and not sentiments, are what drive identity labeling decisions.  This is what we do for the Estimated PCS-FA.

The only question that remains, then, is how to define the values of $\beta$ for Task 2, and how these ideas map to Task 1.  For the Hand-coded PCS-FA, we define $\beta$ manually for both tasks.  For the Estimated PCS-FA, we estimate $\beta$ using a logistic regression model. We describe each approach separately below.

\subsubsection{Hand-coded PCS-FA}

\begin{table}[th]
    \centering
    \begin{tabular}{lllr}
       \textbf{Task} & \textbf{Cue} & \textbf{Identity}  & \textbf{$\beta$ index value (matching function score output)} \\ \hline \hline
        1  & Ethel & grandmother & 4 \\
        1  & Ethel & bully  & 1 \\
        1  & Ethel & college student & 1 \\
        1  & someone & grandmother & 0 \\
        1  & someone & bully  & 0 \\
        1  & someone & college student & 0 \\
        1  & Harold & grandfather & 3 \\
        1  & Harold & bully  & 1 \\
        1  & Harold & college student & 1 \\
        1  & someone & grandfather & 0 \\
        1  & someone & bully  & 0 \\
        1  & someone & college student & 0 \\
        
        1  & Brittany & granddaughter & 2 \\
        1  & Brittany & villain & 1 \\
        1  & Brittany & pharmacist & 1 \\
        1  & someone & granddaughter & 0 \\
        1  & someone & villain & 0 \\
        1  & someone & pharmacist & 0 \\
        1  & Johnny & grandson & 2 \\
        1  & Johnny & villain & 1 \\
        1  & Johnny & pharmacist & 1 \\
        1  & someone & grandson & 0 \\
        1  & someone & villain & 0 \\
        1  & someone & pharmacist & 0 \\ \hline
        
        2  & soccer coach & soccer player & 5 \\
        2  & soccer coach& sports fan & 3 \\
        2  & soccer coach & goon & 1 \\
        2  & soccer coach & shop clerk & 1 \\
        2  & doctor & patient & 5 \\
        2  & doctor & paramedic & 3 \\
        2  & doctor & cousin & 1 \\
        2  & doctor & trespasser & 1 \\
        2  & someone  & soccer player & 0 \\
        2  & someone & sports fan & 0 \\
        2  & someone  & goon & 0 \\
        2  & someone  & shop clerk & 0 \\
        2  & someone & patient & 0 \\
        2  & someone & paramedic & 0 \\
        2  & someone & cousin & 0 \\
        2  & someone & trespasser & 0 \\ \hline \hline
    \end{tabular}
    \caption{Assumed output of the PCS-FA matching function for the given identity with the given cue.  Also given is the relevant Task and the experimental condition }
    \label{tab:pcshs_vals}
\end{table}

Table~\ref{tab:pcshs_vals} provides the value for a given index of $\beta$ that corresponded to the given identity.  In other words, these were our assumed scores from the PCS-FA matching function for each identity.  To compute the probability respondents chose one identity over another, we can simply sub these values back into Equation~\ref{eq:est_beta}.  Note, then, that given the form of Equation~\ref{eq:est_beta}, the primary factor is not the absolute values of the assumed scores, but the relative values for identities appearing in the same question.  Additionally, we note that all values were chosen a-priori, before any prediction was performed.

For Task 1, our hand-coded values for $\beta$ are informed by our manipulation checks. Specifically, we included a manipulation check for the connection of these names to the associated ages by asking respondents how old they expected someone named ``Ethel'', ``Harold'', ``Brittany'' and ``Johnny'' to be. On average, Ethel was expected to be around 73 years old; Harold around 60, Brittany around 29 and Johnny around 30. This suggests that our age manipulation worked much better for Ethel, and to a certain extent Harold, than it did for Johnny and Brittany. From a PCS-FA perspective, the name Ethel most strongly activates the trait ``Old'' (as, presumably, does ``grandmother''), while in contrast, ``Johnny'' does not as strongly activate the youth trait we would expect to be associated with the identity ``grandson''.  We set the index for $\beta$ for these identities accordingly. For Task 2, we had no such manipulation checks. Values were therefore set based on researcher assumptions.

\subsubsection{Estimated PCS-FA}

The critical observation for our Estimated PCS-FA is noting that carried across all survey respondents and all questions for this scenario, Equation~\ref{eq:est_beta} is exactly a logistic regression model where the data is $b-a$ and we estimate the final activation values for each identity answer choice with the vector $\beta$. Therefore, we can estimate the most likely final activation score under the PCS-FA model directly from the data.  We do so for each of the scenarios of Task 2 separately.

For Task 1, we use a similar approach, but additionally incorporate the same knowledge gleaned in the Hand-coded PCS-FA from the manipulation check to further improve the predictions of the model. We assume that there is some $N$ and $\phi$ that describes how the traits implied by the names that we cue respondents with (e.g. ``Ethel'') match each identity given as an answer (intuitively, ``grandmother'' but not ``bully''). We estimate a single $\beta$, which is the difference in activation dependent upon whether or not a trait is cued by a given answer choice. However, instead of a binary matrix for $a$ and $b$, after assigning identities indices, we place the values we identify for $\beta$ in the Hand-coded PCS-FA as the values for $a$ and $b$. Essentially, this provides the model with the knowledge that the strength of the combined gender and age cues is strongest for Ethel, followed by Harold, and then Brittany and Johnny.

\subsection{LCSS} \label{sec:specific_comb}

With no data to base our estimates on, it is  not possible in the present work to empirically characterize the new parameters $\mathcal{Z^*}$, $g^*(f^*)$ and $w^*$.  However, as we will show, reasonable assumptions above the form of $\mathcal{Z^*}$ will lead us to a tractable model we can use to make for the experiment defined in the present work.  Specifically, we will assume that sentiments, traits, and associations are independent with respect to the matching function - that is, for example, the sentimental meanings of $a$ before an event are unrelated to the transient associations  or traits of $a$, $b$ or $o$. Regardless of the form of $g^*(f^*)$, this implies that $\mathcal{Z^*}$ is structured as follows, where $\mathbf{0}_{l}$ specifies a zero vector of length $l$ and, as above $\mathbf{Z_{j}}$ represents the $j$th row of the coefficient $Z$ matrix where we assume that we have values for trait and association dimensions:
\begin{equation}
    \mathcal{Z^*} = \begin{bmatrix} \mathbf{Z_{a_e}} & \mathbf{0}_{|T|}  & \mathbf{0}_{|K|}  \\
                          \mathbf{Z_{a_p}} &  \mathbf{0}_{|T|}  & \mathbf{0}_{|K|} \\
                          \dots & \dots  & \dots \\
                          \mathbf{Z_{o_a}} &  \mathbf{0}_{|T|}  & \mathbf{0}_{|K|}\\
                          \mathbf{0}_{|f|} &  \mathbf{Z_{a_{t_0}}}  & \mathbf{0}_{|K|} \\
                          \dots & \dots  & \dots \\
                          \mathbf{0}_{|f|} &  \mathbf{Z_{a_{t_T}}}  & \mathbf{0}_{|K|} \\
                          \mathbf{0}_{|f|} &  \mathbf{0}_{|T|} & \mathbf{Z_{a_{k_0}}}  \\
                          \dots & \dots  & \dots \\
                          \mathbf{0}_{|f|} &  \mathbf{0}_{|T|} & \mathbf{Z_{a_{k_K}}}  \\
        \end{bmatrix}
\end{equation}

This form of $\mathcal{Z^*}$ allows us to split the deflection equation in Equation~\ref{eq:def_COMB} into three separate and independent deflection computations, as shown in Equation~\ref{eq:def_COMB_2} below, where $deflection_{T}$ and $deflection_{A}$ define the subset of the deflection calculation relevant to traits and associative dimensions, respectively:

\begin{equation} \label{eq:def_COMB_2}
	deflection_{LCSS}(f^*) =   \frac{w_{f}}{|f|} deflection(f) + \frac{w_{f_t}}{|f_t|}deflection_{T}(f_t) + \frac{w_{f_k}}{|f_k|}deflection_{K}(f_k)
\end{equation}

Equation~\ref{eq:def_COMB_2} shows that while we can express LCSS as a single, intuitive theoretical construct, we can at the same time, with a reasonable assumption, decouple the three different components of the model for use in an identity labeling context. This is convenient for the present work, as if we are able to define reasonable assumed values for $deflection_{T}(f_t)$ and $deflection_{K}(f_k)$ for questions in our vignette, we can then  leverage a similar logistic regression framework as in Estimated PCS-FA to estimate the relative weights of sentiment, trait, and association deflection in computing $deflection_{LCSS}$. 

We first explain how we characterize associative deflection, $deflection_{K}(f_k)$, for the various vignette questions. Values are hard-coded using the assumptions of Hand-coded PCS-FA, where we assume a score for each Actor and Object pairing as described in Table~\ref{tab:pcshs_vals}. The only exception to this is that for the scores output by the PCS-FA matching function, a higher score means a stronger association. For LCSS, $deflection_{K}(f_k)$ should be high when we do \emph{not} expect the actor and object to be seen together (i.e. the transient associative meanings would have to shift for this event to make sense to us).  Consequently, rather than using the values for $\beta$ in Table~\ref{tab:pcshs_vals}, we first find the maximum value for the assumed matching score output in Task 2 (that value is 5), and then subtract the corresponding value for the given cue/identity pair to compute $deflection_{K}(f_k)$. So, for example, the role pairs soccer coach and soccer player have an associative deflection of 0, implying that our associative meanings do not change between the fundamental and the transient for these identities, because we already expect to see these identities together. Where unspecified, we set $deflection_{K}(f_k) = 0$. Specifically, in Task 1, we assume that no associative deflection occurs, because associative meanings are not specified.  Alternatively, this is the same as assuming that associative deflection is the same for the two choices.  Trait deflection, $deflection_{T}(f_t)$, is specified in an analogous way for Task 1 and Task 2.

Having specified values for $deflection_{K}(f_k)$ and  $deflection_{T}(f_t)$, we can complete the steps necessary to constructing a $\phi$ for LCSS by using the negation of the deflection equation in Equation~\ref{eq:def_COMB_2}. This is analogous to the approach we take for ACT in Equation~\ref{eq:act_phi}, for brevity we do not repeat it here.  Therefore, we have the following for LCSS for our survey questions, where we use $d$ as a shorthand for $deflection$, e.g. $deflection_{LCSS} = d_{LCSS}$, and $f^*_{i_a}$ is the fundamental vector for a given question with $i_a$'s values
\begin{align}\label{eq:COMB_phi}
    \phi(i_y = i_a) &= \frac{1}  {1 + \exp \left(\phi(i_b,M,r)-\phi(i_a,M,r)\right)} = \frac{1}  {1 + \exp \left(d_{LCSS}(f^*_{i_a}) - d_{LCSS}(f^*_{i_a})\right)} \nonumber \\
    &= \frac{1}  {1 + \exp \left( \frac{w_{f}}{|f|} d(f_{i_a}) + \frac{w_{f_t}}{|f_t|}d_{T}(f_{t,i_a}) + \frac{w_{f_k}}{|f_k|}d_{K}(f_{k,i_a}) -     \frac{w_{f}}{|f|} d(f_{i_b}) - \frac{w_{f_t}}{|f_t|}d_{T}(f_{t,i_b}) - \frac{w_{f_k}}{|f_k|}d_{K}(f_{k,i_b}) \right)} \nonumber \\
    & = \frac{1}  {1 + \exp \left( \frac{w_{f}}{|f|}( d(f_{i_a}) - d(f_{i_b})) + \frac{w_{f_t}}{|f_t|}(d_{T}(f_{t,i_a}) - d_{T}(f_{t,i_b} )) + \frac{w_{f_k}}{|f_k|}(d_{K}(f_{k,i_a}) - d_{K}(f_{k,i_b})) \right)}
\end{align}

The only unknown values in Equation~\ref{eq:COMB_phi} are the weights on each of the separate forms of deflection. As in Equation~\ref{eq:est_beta}, we can therefore estimate these weights from survey data, giving an indication of how important each of these factors are in predicting how individuals will label others. 

\section{Results}\label{sec:results}

\begin{figure}[t]
    \centering
    \includegraphics[width=.95\textwidth]{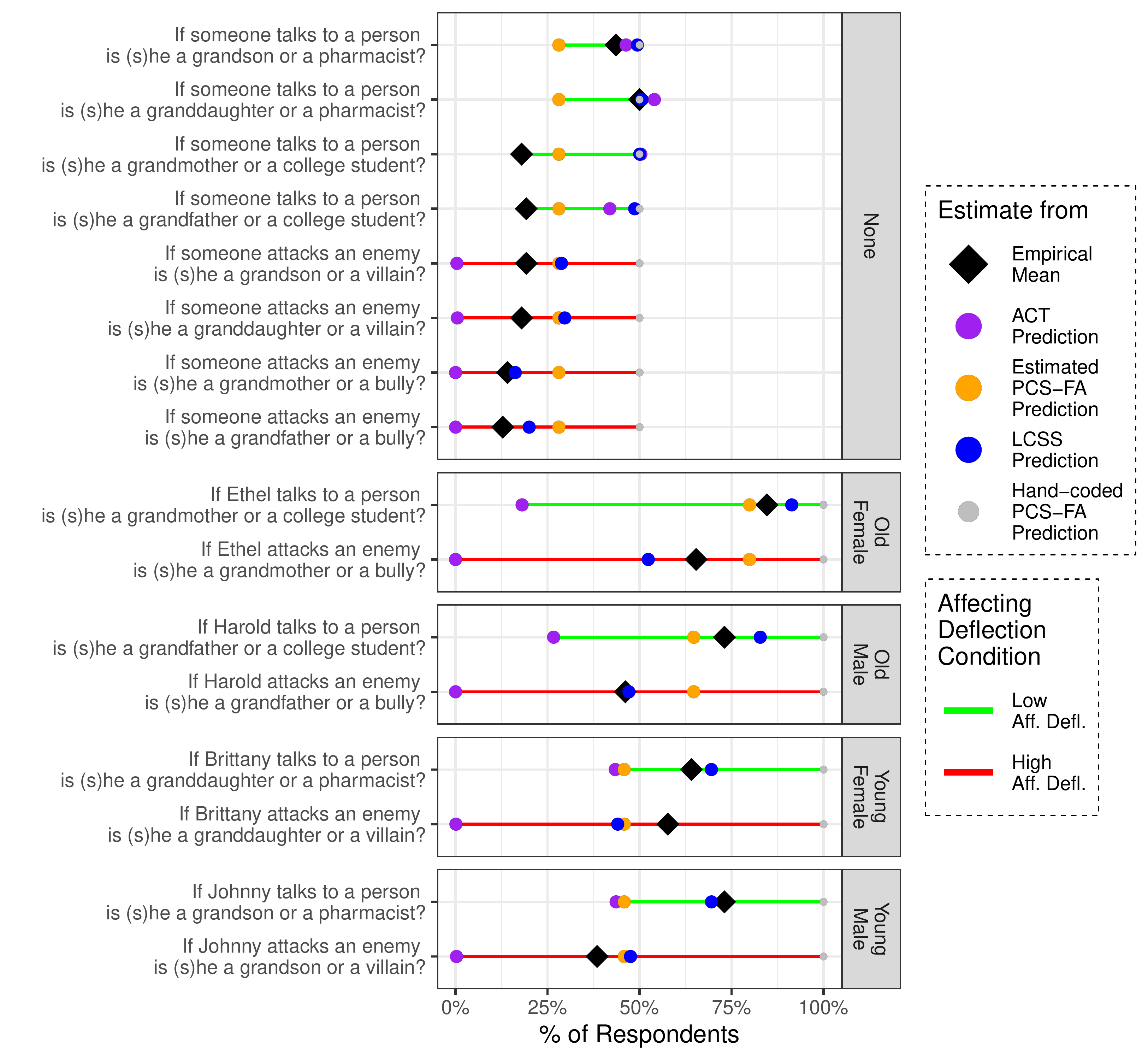}
    \caption{On the y-axis, the text of each vignette in Task 1. The x-axis gives the percentage of respondents who chose the first answer given in the question. Grey subplot labels separate trait conditions, line color distinguishes sentiment conditions. The different color points represent estimates of this value for the four identity labeling models (Purple circles for ACT, Orange circles for Estimated PCS-FA, Grey for the Hand-coded PCS-FA, Blue for LCSS) and the empirical prediction (black diamonds).}
    \label{fig:full_study1_task1}
\end{figure}

\begin{figure}[t]
\centering
   \includegraphics[width=.85\linewidth]{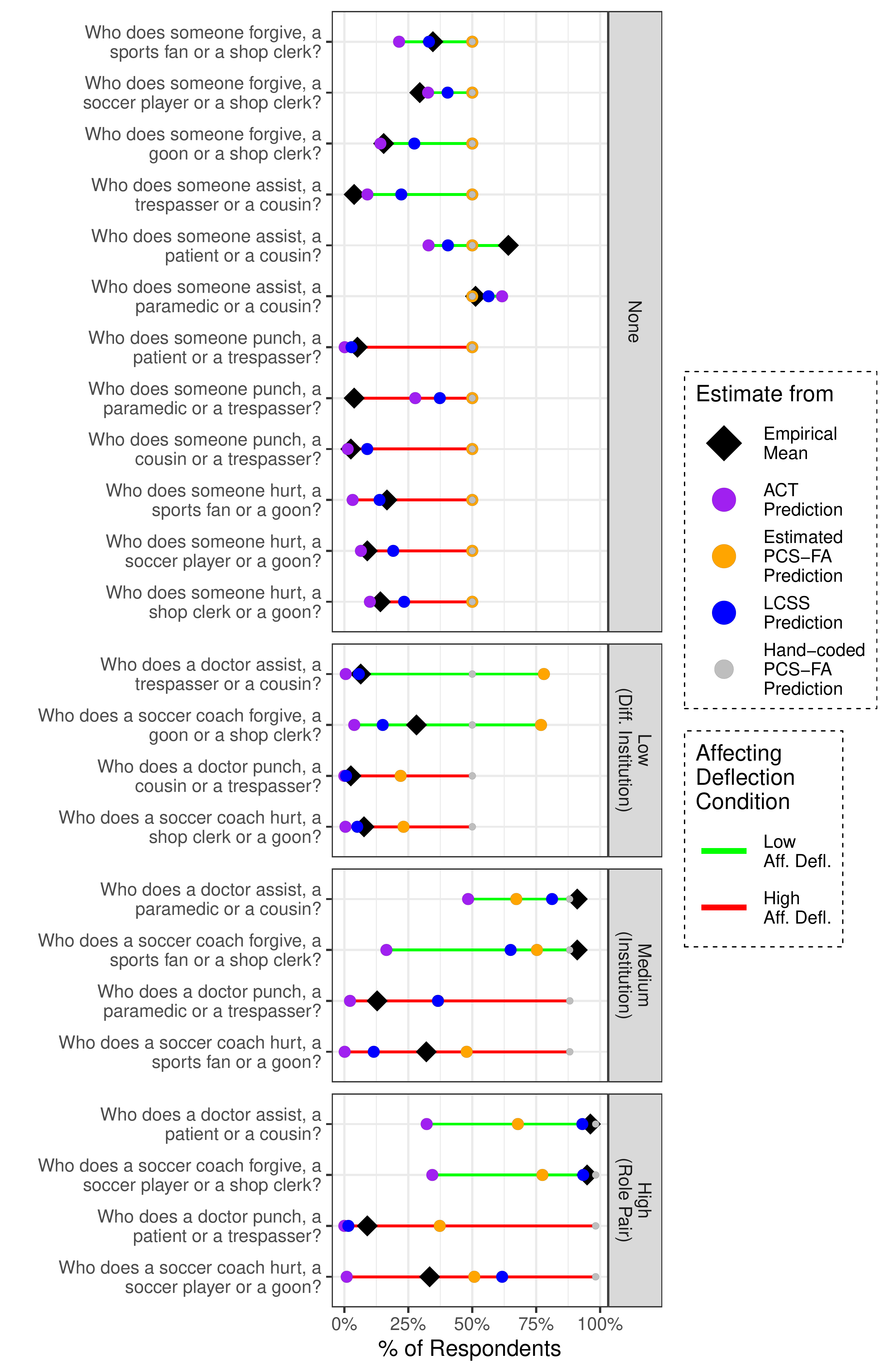}
\caption{On the y-axis, the text of each vignette in Task 2. Grey subplot labels separate association conditions, line color distinguishes sentiment conditions. The x-axis gives the percentage of respondents who chose the first answer given in the question (e.g. ``soccer player'' in the top question on the y-axis). The different color points represent estimates of this value for the four identity labeling models (Purple circles for ACT, Orange circles for Estimated PCS-FA, Grey for the Hand-coded PCS-FA, Blue for LCSS). The black diamond gives the empirical result.}
\label{fig:study1_task2}
\vspace{-1.2em}
\end{figure}

\begin{figure}[t]
    \centering
    \includegraphics[width=\linewidth]{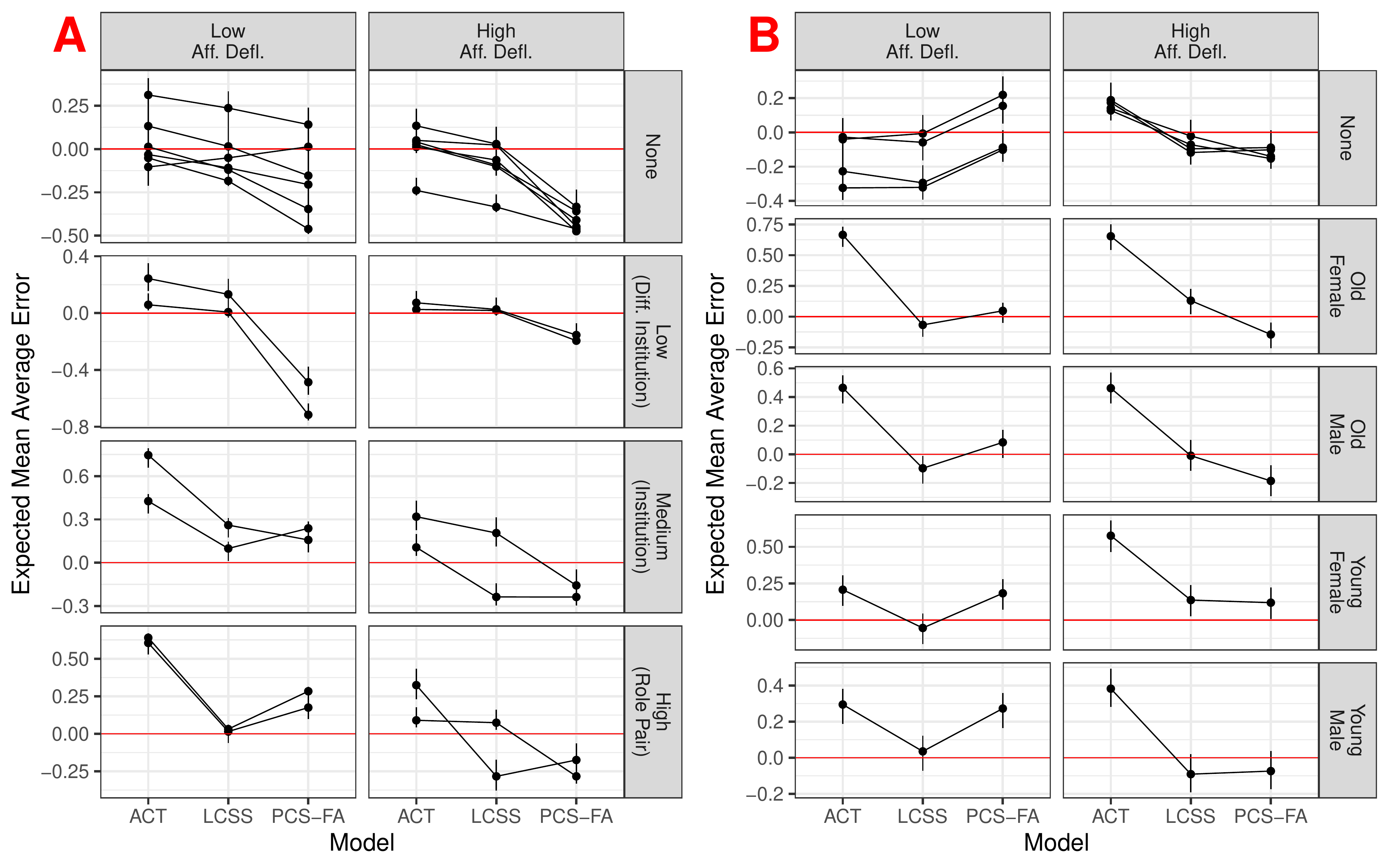}
    \caption{a) (Left), Prediction errors for Task 1  for ACT, LCSS, and Estimated PCS-FA. b) (Right), The same, but for Task 2 (right).  On the x-axis of both figures, the models are listed.  The y-axes gives the expected error of that model, i.e. the difference between the predicted percentage of respondents who selected Answer 1 and the actual percentage. Estimates include 95\% confidence intervals, computed using the Agresti-Coulli method \citep{agresti_approximate_1998}. Each line represents a different question asked in Table~\ref{tab:study1_task1}. Each subplot presents results for a different affective condition, labeled by the grey bar differentiating the left and right columns of plots, and a different trait or association condition (the different rows).}
    \label{fig:study1_errs}
\end{figure}

\subsection{Main Findings}

Results show that PCS-FA and ACT outperform each other in theoretically consistent ways, and that LCSS is a better predictor than either PCS-FA or ACT. PCS-FA should perform better than ACT when associations and/or traits are present, but where there is minimal affective deflection. For example, we would expect the PCS-FA model to better predict responses to the question ``Who does a doctor assist, a patient or a cousin?'' than ACT, because we expect respondents to leverage the association between doctor and patient and assume they rely less on the limited differences in affective deflection between the two options. Indeed, on the eight vignettes in the low (difference in) sentiment condition where traits (Figure~\ref{fig:full_study1_task1}, bottom four panels and green bars)  or medium-to-high associations (Figure~\ref{fig:study1_task2}, bottom two panels, green bars) are cued , the MAE of ACT is 50.6\%— around three times that of the Estimated PCS-FA (18.0\%) or Hand-coded PCS-FA (14.5\%). 

Differences are statistically significant (p \textless .01). ACT should perform better than PCS-FA when no information about traits or associations is provided. For example, in Task 1, we expect ACT should out-perform PCS-FA in predicting responses to the question ``If someonee attacks an enemy, is (s)he a grandson or a villan,'' because respondents have almost no semantic information, and a clear choice in terms of affective deflection. Indeed, in these conditions, the MAE for ACT was only 11.9\%— nearly two times better than the MAE of the Estimated PCS-FA (24.2\%) or Hand-coded PCS-FA (29.3\%) (p \textless .01). In addition to the provided statistics, these improvements can be seen in Figure~\ref{fig:full_study1_task1}, which presents results for all questions and models for Task 1, and Figure~\ref{fig:study1_task2}, which presents the same for Task 2.

Turning to the combined model, the MAE of LCSS across all questions is 10.9\%, with a 95\% bootstrapped confidence interval of [8.2\%-14.2\%]. ACT (24.3\% [18.3\%-31.5\%]), Estimated PCS-FA (23.0\% [18.7\%-28.0\%]), and Hand-coded PCS-FA (33.4\% [26.9\%-40.0\%]) all perform significantly worse. In absolute terms, the combined model out-predicts the other two models by over 100\%.  This improvement is not only the result of averaging when sentiments, and traits or associations, are present. As Figure~\ref{fig:study1_task2} shows, LCSS is also better than PCS-FA when only affective signals are provided, and no trait or association information, e.g. in the ``None'' condition in the top panel of Figure~\ref{fig:full_study1_task1}. It is also better than ACT under conditions of limited affective deflection, where trait and association signals instead dominate, as in the low affective deflection (green bars) in the bottom four sub-panels of Figure~\ref{fig:full_study1_task1}. A combined model is therefore useful to address not only cases where signals must be averaged, but also cases where signals of particular dimensions are absent on certain dimensions of meaning, without requiring a-priori assumptions on when signals are present or absent.


\subsection{Error Analysis}

While it is instructive to find that the combined model out-performs either ACT or PCS-FA on their own, we can learn more about why this is the case by digging into the errors of each model. To this end, we provide Figures~\ref{fig:study1_errs}a) and b), which provides a more complete view of the errors of each model.  These figures emphasize the claim that a combined model performs better than the other two models in part because it improves poor predictions from ACT when affective information does not differentiate between the two options but trait/association information does. This is most clear on the bottom two subplots in the right side of Figure~\ref{fig:study1_errs}a). Because the combined model weights both affective and trait/association information, it can, unlike ACT, account for cases when semantic information is prevalent and, thus, use it to predict respondent answers. This is the case in the low difference in affective deflection conditions.  In our implementation of PCS-FA, when no trait information is provided, PCS-FA models assume the two answers are equally likely, because they do not incorporate any semantic information. In contrast, the combined model captures sentiments and can therefore characterize differences between the two identities. Although the combined model has several more parameters than the PCS-FA model, it is notable that the combined model improves performance over PCS-FA models while \emph{estimating from the data} fewer parameters (Task 1) or only one additional parameter (Task 2).

Figures~\ref{fig:study1_errs}a) and b) also emphasize, however, that the combined model does not provide perfect predictions. Several of its predictions are outside of the 95\% confidence interval from the survey data. Additionally, improvement from the combined model derives mostly from cases where ACT leans heavily on limited affective differences  (the low affective deflection conditions) or where PCS-FA has limited or no cognitive information to draw on (the No Trait condition in Task 1, and the None condition in Task 2). These two cases suggest that the combined model helps to ameliorate issues where one model simply does not have the correct information to draw from. Where both signals exist, however, the combined model does not obviously improve upon predictions of the two models. For example, the combined model does not accurately predict survey responses for the question ``Who does a soccer coach hurt, a soccer player or a goon?'' This is likely because the current stimuli does not allow the combined model to learn which information- affective or semantic- to weigh more (in this question, for example respondents seemed to rely more heavily on affective information) and when to do so.   

Finally, note that ACT does not always perform well even when no trait or associations are provided. For example, the theory poorly predicts responses to the question, ``If someone attacks an enemy, is (s) a grandson or a villain?'' This is because the quality of ACT’s prediction, and thus its contribution to the combined model, relies heavily upon the affective coherence of the event being modeled. In numerous cultures over multiple time periods, scholars have observed interaction effects that relate to the expectation of situational consistency \citep{smith-lovin_affect_1988,morgan_distinguishing_2016}. The expectation that bad people do bad things to other bad people is a common example. There are several interactions of this kind in the theory’s impression change equations. Consequently, when a situation involves a combination of identities and behaviors that are affectively inconsistent in terms of these expectations, the model’s predictions becomes less informative. The fact that the model struggles with these kinds of events is, from our perspective, a reflection of the influence of prevailing affective expectations that regulate a culture’s social interactions. Humans confronted with these situations in vignette studies often note frustration and confusion with vignettes that violate these expectations \citep{rogers_you_2018}.  Nevertheless, the linkages between situational coherence and deflection requires formal specification. Presently, we do not model the impact that coherence has on deflection. We suggest incorporating variance components to capture the effect of situational coherence, at the very least to provide a measure of our confidence in the model’s predictions.

\section{Discussion and Conclusion}

The present work formalizes the identity labeling problem, compares predictions of  two existing mathematical models that can be used to address it, and introduces a new modeling framework, LCSS, that combines these existing models. Overall, we find that the two primary existing models of identity labeling---affect control theory and the parallel constraint satisfaction model--- perform well on vignettes that they are designed for but struggle when information they expect is not cued, and when presented with cues for dimensions of meaning they are ill-suited to accommodate. We combine the complimentary strengths of these two models into a new approach, that we call Latent Cognitive Social Spaces. We show that using this model, we can partially address limitations of existing approaches, avoiding problems due to incomplete information and problems that arise when all signals are present simultaneously. This is because ACT and PCS-FA focus on complementary sets of signals that humans use to interpret social situations. Combined, these models provide predictive power from theoretically and conceptually distinct dimensions of meaning. 

The present work is somewhat unusual in that our primary goal is to introduce a mathematical framework for testing hypotheses related to identity and identity maintenance rather than testing a particular hypothesis. LCSS is motivated, first and foremost, by the observation that a) existing models for a given problem are likely to be poor predictors in certain situations and that b) these situations are complementary.  We theorize that the concept of deflection when conceptualized as a signal of situational tension can encompass multiple kinds of cues about the violation of cultural expectations, and that individuals use this more general form of deflection when labeling others.  In this paper, we introduce a framework necessary for exploring these claims by adapting affect control theory's mathematical structure. Because our goal was to provide a framework rather than to answer a specific theoretical question, numerous questions remain as to how to move forward with LCSS, both theoretically and mathematically.

Three questions in particular seem crucial to address in subsequent work. First, LCSS relies on the idea that deflection is a meaningful theoretical construct for traits and semantic associations. \citeauthor{mackinnon2020inextricability}'s (\citeyear{mackinnon2020inextricability}) and \citeauthor{RN8}'s (\citeyear{RN8}) work on the management of denotative and connotative meanings provides an alternative but complimentary approach to understanding how these two forms of meaning play a role in identity labeling. Borrowing from that approach, leveraging (un)certainty about the relationship between traits and affective meanings in the matching function would likely improve the predictive power of LCSS. Nevertheless, there are still several exciting but thorny theoretical questions regarding what situational cues people are likely to focus on in different situations. Second, the general LCSS model assumes an unknown number of meaningful dimensions of traits and associations.  We are able to avoid addressing this in our empirical work due to a careful experimental design, but we need to better understand both what those dimensions are and their consistency across cultures.  We believe that efforts to do so that combine experimental data with text analysis are likely to be particularly fruitful \cite{keith2020text}. Finally, the number of parameters in the general LCSS model is large; specifically, it is dependent on the square of the number of assumed traits, associations, and sentiments that are socially meaningful. Again, our experimental design allows us to side-step these issues in the present empirical work. However, how to best estimate theoretically meaningful subsets of these parameters is a critical avenue for future investigation.

These questions are particularly interesting in that LCSS, while designed here as a way to address the identity labeling problem, is based largely on the principle of affect control which is much more general. Another perspective on our work, then, is to show that perhaps the most popular and well-tested mathematical model of social behavior, affect control theory, can be extended to incorporate cognitive and affective signals in combination to predict how individuals make sense of and act in social situations. Our work, thus, opens several new theoretical and empirical doors in the mathematical analysis of social interaction.

\bibliographystyle{chicago}
\bibliography{pnas-sample}

\appendix

\section{BayesACT as an Identity Labeling Model}\label{app:bayesact}
 that we have not addressed its successor, BayesACT.  In short,  BayesACT assumes that the deflection equation in Equation~\ref{eq:bayesact} is the logarithm of a probabilistic potential function \citep[Equation 7 in ][]{hoey_bayesian_2013}):
\begin{align}\label{eq:bayesact}
	q(f',\tau') \propto \exp\left(-(f'-\tau')^T\Sigma^{-1}(f'-\tau')\right) 
\end{align}

In this model, the fundamental is defined as $f'$ and the transient is defined as $\tau'$.  This is done in order to emphasize that in BayesACT, unlike in ACT, the fundamental and the transient are (time varying) probability distributions. Because it assumes a probabilistic form, BayesACT, thus, can directly give a probability that a given identity is the ``correct'' identity for a given situation.  Relevant to the present work is that the form of Equation~\ref{eq:bayesact} is clearly similar to the form of Equation~\ref{eq:choice}; indeed, Equation~\ref{eq:choice} is simply a normalized version of the model posed by \cite{hoey_bayesian_2013} where we also integrate over $f'$ and $\tau'$:
\begin{align}
	p(i_y | M, r) = p(i_y | f',\tau') \propto \frac{ \int_{f',\tau'} q(f_{i_y}',\tau_{i_y}')}{\sum_{k \in I}  \int_{f',\tau'} q(f_{i_y}',\tau_{i_y}')} 
\end{align}

This is a discrete choice model in the form of Equation~\ref{eq:choice} where $\phi(i_a, M,r) = \int_{f',\tau'} \log(q(f_{i_y}',\tau_{i_y}'))$. In Appendix B of \citep{hoey_bayesian_2013}, the authors demonstrate how BayesACT reduces to the same predictions as the original ACT model under certain conditions. Because computing the integration over $f'$ and $\tau'$ by either derivational or numerical approaches would complicate the simplicity of the mathematical argument posed in the present work, we will, therefore, assume that it is under these conditions that the predictions are made here. In other words, we use predictions from the ``original'' version of ACT described above, leaving a comparison with predictions made by BayesACT under different parameter settings to future work.

One final point worth noting regarding BayesACT is that the theory, unlike ACT, requires no assumption that identities are known for social interaction to take place.  Instead, BayesACT only requires a \emph{distribution} over identities is known for the actor and object.  The present work, instead, focuses on data and methods that imply people often choose a label for actors and object. It should be noted, however, that the more general definition of the identity labeling problem proposed in Section~\ref{app_sec:formal} requires only this same probability distribution considered by BayesACT; the final selection is necessary to compare with our survey results but not in principle.  In this way, the formalizations of the identity labeling problem are exactly the same as those embraced by BayesACT.

\end{document}